# Role of Ligand Conformation on Nanoparticle-Protein Interactions

*Federica Simonelli[1], Giulia Rossi[1]\* and Luca Monticelli[2]\**

[1] Physics Department, University of Genoa, Via Dodecaneso 33, 16146 Genoa, Italy
[2] MMSB, UMR 5086 CNRS / Universitè de Lyon, 7, Passage du Vercors, 69007 Lyon, France

**Corresponding Authors**

*rossig@fisica.unige.it, luca.monticelli@inserm.fr

ABSTRACT

Engineered biomedical nanoparticles (NP) administered via intravenous routes are prone to associate to serum proteins. The protein corona can mask the NP surface functionalization and hamper the delivery of the NP to its biological target. The design of corona-free NPs relies on our understanding of the chemical-physical features of the NP surface driving the interaction with serum proteins. Here we address, by computational means, the interaction between human serum albumin (HSA) and a prototypical monolayer-protected Au nanoparticle. We show that both the chemical composition (charge, hydrophobicity) and the conformational preferences of the ligands decorating the NP surface affect the NP propensity to bind HSA.

INTRODUCTION

Nanoparticles designed to be administered via intravenous routes are prone to interact with serum proteins, which can stably cluster around the nanoparticle forming a protein corona[1–4]. The non-specific adsorption of proteins on NPs alters their designed function and influence their fate in the body[5–7]. The control of protein adsorption[8–10] and the minimization of early clearance from the bloodstream are crucial to the clinical integration of synthetic nanoparticles[6]. Most often, inorganic NPs designed for diagnostic or therapeutic applications do not expose their bare surface to the biological environment, but are functionalized by organic ligands that provide better solubility and specific targeting properties. The density, length, charge, and hydrophobicity of the NP ligands determine the amount and type of proteins that bind to the NP[8,11,12], as well as the reversibility of binding[13–16].

One possible route to act on the NP-protein interaction, in the direction of reducing non-specific adsorption, involves the functionalization of the NPs with proper anti-fouling functional groups. Polyethylene glycol (PEG) is known to be a good anti-fouling material[17], and a consistent body of literature has shed light on its action as a stealth agent. Protein-repellent properties of PEG grafted on surfaces are influenced by PEG chain length[18], density, and environment temperature[19,20], though not always the amount of adsorbed proteins is a monotonic function of these parameters[21]. The use of PEG as a stealth agent also has some drawbacks, such as its non-biodegradability, immunogenicity[22], and its accumulation in membrane-bound organelles[23]. An alternative to PEG is represented by ligands terminated by zwitterionic moieties, which further reduce non-specific protein adsorption[13,22,24]. Zwitterionic groups can thus extend the circulation time of the NPs and increase their ability to effectively penetrate cell membranes[5].

The many physical and chemical parameters that characterize the NP-protein interface, on both sides, make it difficult to identify clear correlations between the composition of the NP surface and the composition and stability of the protein corona. The computational approach can contribute to shed light on which factors, on the molecular scale, determine the formation of stable coronas. Molecular simulations face important limitations, though, as the corona formation is a process that spans time scales of seconds, the relevant NP sizes for biomedical applications range from a few to hundreds of nm, while the thickness of the protein corona on metal or metal oxide NPs varies from 20 to 40 nm[8]. The simulation, at atomistic or molecularly detailed coarse-grained resolution and with explicit solvent, of a whole NP+corona complex is still out of reach for current computational resources. The use of implicit solvent schemes has allowed for the simulation of the corona formation on top of model spherical NPs[25,26]. Several attempts have been made at the simulation of the interaction of a single NP with a single protein. This has most often required to give away significant details of the NP-protein interface. NPs are often modeled as flat surfaces[27] or smooth spherical objects, offering a generic hydrophobic, hydrophilic or charged surface to the protein[28–30]. Proteins, as well, may be treated as rigid bodies[29] or polymers with no secondary structure[30].

Here we use Molecular Dynamics, at coarse-grained (CG) resolution, to investigate the interplay of electrostatics, hydrophobicity, and ligand conformation at shaping NP-protein interactions. Our model combines an atomistic description of the Au core[31] to a coarse-grained, explicit solvent model of the rest of the system. The coarse-grained description has sub-molecular resolution, and it takes into account explicitly the composition of the NP ligand shell, its flexibility, and protein flexibility. We simulate the interaction between human serum albumin (HSA) and monolayer-protected Au NPs. HSA accounts for more than half of the serum proteins in human blood plasma[32], and it is one of the most abundant components of the corona formed around nanoparticles[11,12], and specifically ligand-protected Au NPs[14]. The Au NPs we consider have the same composition and size of those synthesized by Moyano *et al*[13]. The Au core has a small diameter of 2 nm (4 nm in the experiments by Moyano *et al*[13]). The Au surface is covalently functionalized by ligands which are terminated by a zwitterionic group and, at the same time, have tunable hydrophobicity. This ligand composition offers the opportunity to monitor the influence of electrostatic and hydrophobic interactions at shaping the NP-protein interaction.

METHODS

As a first step, we developed a coarse-grained model of HSA in the framework of the polarizable-water Martini coarse-grained model, which allows realistic large-scale simulations of proteins[33–35] and nanoparticles[36,37]. The Martini force field does not allow for changes of the protein secondary structure, which is imposed by means of an elastic network connecting the CG beads that are placed on top of the $C_\alpha$ atoms[38]. This description of HSA is compatible with the indication that the secondary structure of HSA does not change upon binding to Au surfaces[27] and nanoparticles such as fullerenes[39] and Au nanoparticles in the 4-40 nm range[14,40]. The development of the CG model was based on structural and dynamic parameters obtained from atomistic simulations carried out with the Amberff99SB-ILDN force field[41]. Several CG models were tested, with different parameters defining the elastic network. We considered two structural parameters (the root-mean-square deviation of the alpha carbon atoms, RMSD, and the per-residue root-mean-square fluctuation, RMSF). We also performed principal component analysis (PCA) to quantify the superposition of principal components (PCs) in atomistic and CG simulations. Finally, we selected the CG model most similar to the all-atom model in terms of RMSD and RMSF, and with the highest overlap in PCA. All the details about the model development are reported in the Supporting Information.

In order to test the influence of hydrophobicity on the interaction between zwitterionic NPs and HSA, we tested two different NP models. The two NPs have an identical core of 144 Au atoms and differ only for the composition of their 60 ligands (Figure 1). The least hydrophobic NP, referred to as Z, has ligands composed by a short hydrophobic stretch, a sequence of 4 monomers of PEG and a zwitterionic sulfobetaine terminal. The most hydrophobic NP, referred to as ZH, has identical ligands except for two additional hydrophobic branches stemming from the zwitterionic terminal group. The details of ligand parameterization are reported in the Supporting Information.

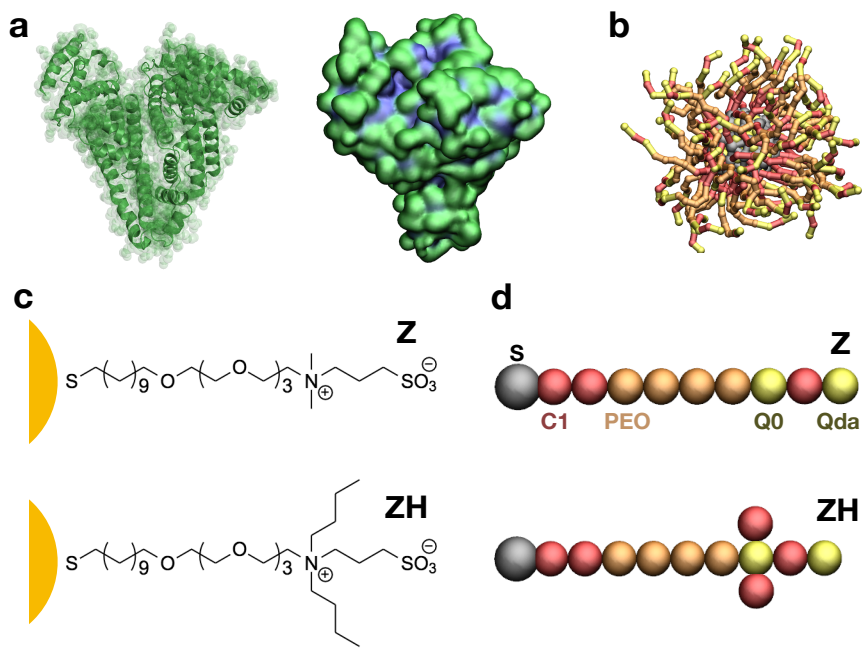

**Figure 1** – a. HSA. On the left, the secondary structure of the protein; on the right, the protein surface colored according to hydrophobicity (hydrophobic residues in blue, and charged or polar residues in green); b. The ligand-protected NP, with grey Au core and S atoms, and Z ligands. c. Chemical composition of the Z and ZH ligands of the NP. d. The Z and ZH ligands as represented by the CG model; C1, Q0 and Qda refer to non-bonded types of the Martini force field[42], PEO is the Martini type defined in Lee *et al.*[43]

RESULTS AND DISCUSSION

We characterized the NP-HSA interaction by means of unbiased MD runs in which a single NP was allowed to interact with a single HSA protein. We performed 20 runs at physiological conditions (310 K, atmospheric pressure) with a time step of 20 ns for a total simulated time of 60 microseconds for each NP type. All simulations were run with the GROMACS 5 package. More details on the MD settings can be found in the Supporting Information. Both NPs are found to establish transient contacts with the protein. To quantify the number and temporal stability of NP-protein contacts, we consider the NP and the protein to be in contact when at least two of their CG beads are closer than a threshold distance of 0.8 nm. The ZH NP resides on HSA surface for longer stretches of time (see Figure 2) compared to the Z NP. For the ZH NP, the total time spent in the bound state is $t_{ZH}^b$ = 23.8 μs over the simulated $t_{run}$ = 60 μs, while for the Z NP $t_Z^b$ = 4.9 μs over the same $t_{run}$. The free energy difference between the bound and unbound state can be thus estimated as $\Delta G_{ZH} = -RT\ln\left(\frac{t_{ZH}^b}{t_{run}-t_{ZH}^b}\right)$ = 1.1 kJ/mol (0.43 $k_B$T), and $\Delta G_Z = -RT\ln\left(\frac{t_Z^b}{t_{run}-t_Z^b}\right)$ = 6.2 kJ/mol (2.4 $k_B$T). We remark that these energy differences do not refer to the binding of the NP to a specific site, but take effectively into account all the binding mechanisms observed during the simulations.

As both NPs undergo many binding and unbinding events during the simulation time, it is also possible to extract information about the effective free energy barriers for binding and unbinding. We define the average residence time as the average time duration of a binding event. The average residence time of the ZH NP is $\langle t_{ZH} \rangle = 34.2 \pm 0.3$ ns, while for the Z NP is $\langle t_Z \rangle = 3.91 \pm 0.01$ ns. Based on the mean residence time, we can estimate the difference, $\Delta^{u\ddagger}$, between the effective unbinding free energy barriers for the two NP types:

$$\frac{\langle t_Z \rangle}{\langle t_{ZH} \rangle} = e^{-(\Delta G_{ZH}^{u\ddagger} - \Delta G_Z^{u\ddagger})/k_B T}$$

$$\Delta^{u\ddagger} = \Delta G_{ZH}^{u\ddagger} - \Delta G_Z^{u\ddagger} = -k_B T \ln(\langle t_Z \rangle / \langle t_{ZH} \rangle)$$

where we have indicated with $\Delta G_i^{u\ddagger}$ the height of the unbinding barrier for the NP of type *i*. $\Delta^{u\ddagger}$ results to be equal to 2.25 $k_B$T. As for binding, the average time spent by the two NPs in the unbound state, that is in the water phase, $\langle t^w \rangle$, is similar: $\langle t_{ZH}^w \rangle = 51.6 \pm 0.2$ ns, and $\langle t_Z^w \rangle = 42.9 \pm 0.1$ ns, corresponding to a difference of 0.19 $k_B$T between the binding free energy barriers. Figure 2 shows, in the bottom panel, a sketch of the free energy of the bound, unbound and transition states for the two NPs.

In order to further probe the scarce propensity of the zwitterionic NPs to stably bind HSA, we also performed a comparison with a NP functionalized by PEG ligands, with the same density and length of the Z and ZH ligands. With PEG ligands, we found that the NP-HSA binding is irreversible on the simulation time scale (3 microseconds); the result is robust against the use of different PEG parameterizations[43,44] (see Fig. S1 and the Supporting Information for a detailed description of these simulations). These results are in excellent agreement with the experimental findings by Moyano *et al.*[13], suggesting that no hard corona is formed on the surface of 4 nm Au

NPs with a zwitterionic ligand shell, while it is formed on NPs functionalized by neutral PEG ligands[13]. Moreover, the small difference in the energy barrier for the unbinding of the Z and ZH NPs is consistent with the small, reversible precipitation observed in the experiments for the most hydrophobic NPs[13].

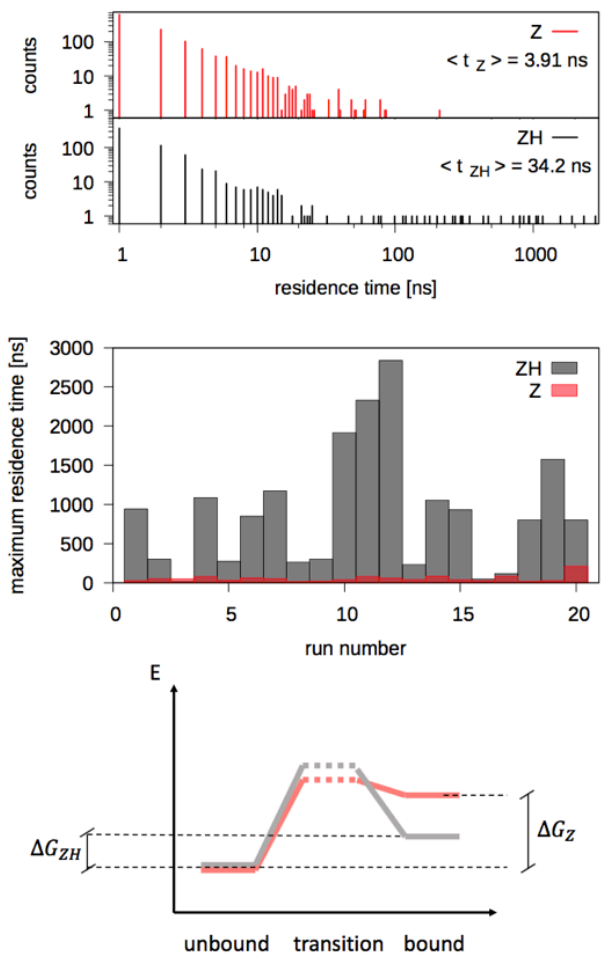

**Figure 2** Top: distribution of NP-protein residence times for Z and ZH NPs. Center: maximum residence time during each of the 20 unbiased MD runs, for each NP type. Bottom: sketch of the free energy barriers for binding and unbinding of ZH and Z NPs (same color code as above)

The different residence times of the Z and ZH NPs suggest that the increased binding of ZH is due to the contribution of the additional hydrophobic groups on the ZH surface. To verify this hypothesis, we analyzed in more detail the nature of the contacts between the protein and the two NPs. **Figure 3** shows that the binding of the Z NP to HSA is quite uniform on the protein surface, while two preferential binding sites emerge from the interaction between ZH and HSA. These binding sites have different shapes (one has the form of a protrusion, the other one of a pocket) and contain both hydrophobic and charged residues. We classified the contacts between HSA and the NPs as hydrophobic, charged, or polar, depending on the character of the amino acid involved (details on the classification can be found in the Supporting Information). Surprisingly, hydrophobic contacts are roughly the same for Z-HSA and ZH-HSA binding (Figure 3c).

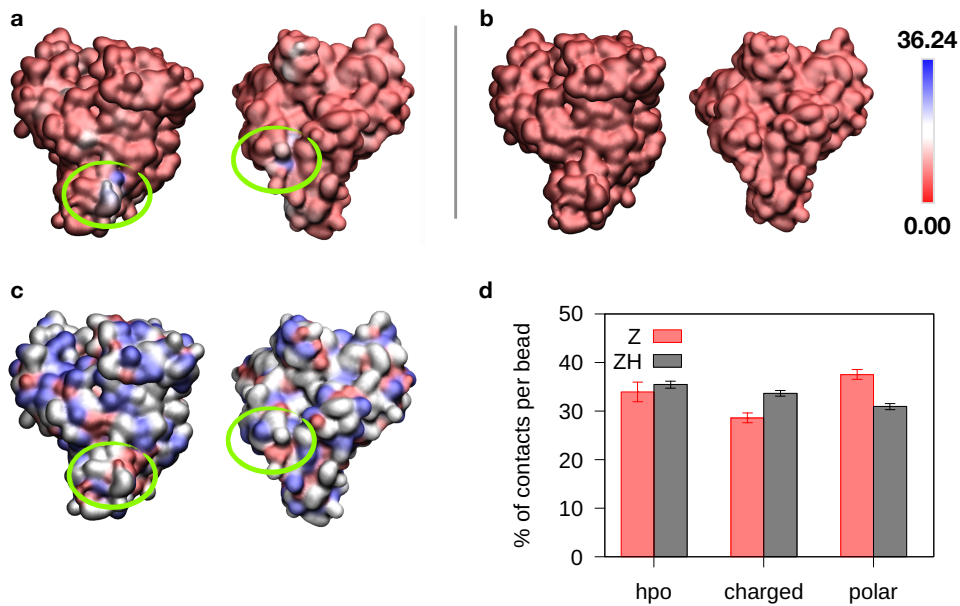

**Figure 3** a. Protein surface colored based on the average number of contacts with the ZH NP. b. Same colormap, for the Z NP. c. Protein surface colored based on residue polarity (hydrophobic residues in red, charged in white, polar in blue). d. Histogram of protein-NP contacts involving hydrophobic, charged and polar residues of HSA.

Even more surprising is the picture emerging from the classification of the NP-HSA contacts based on the type of group of the NP ligand bound to the protein, as shown in **Figure 4**. The main difference between the two NP types is represented by the number of contacts established by the PEG segment of the ligand, significantly higher for the ZH NP. Unexpectedly, the number of hydrophobic contacts is lower for the ZH NP. Why do more hydrophobic ligands make less hydrophobic contacts with the protein? The answer is provided by the radial distribution functions of the different groups composing the NP ligands, as shown in the bottom panel of **Figure 4**. The terminal groups of the Z NP ligands reach out for the water phase, indicating that the ligands mainly have an extended conformation. On the contrary, the (more hydrophobic) terminal groups of the ZH ligands are found closer to the Au surface, well screened from interactions with water, indicating that the ligands mainly have a folded conformation. Such folding brings the central PEG segment of the ligand chains to the water interface, promoting PEG-HSA contacts. These data highlight that another important physical parameter affects NP-protein interaction: ligand conformation.

The slight difference between the free energy barriers for binding observed for the Z and ZH NPs can also be interpreted as a consequence of the different ligand conformations. Indeed, the conformational change induced by the presence of the $C1_T$ groups also affects the hydration of the NP. The bottom panel of **Figure 4** shows the radial distribution function of water beads (PW) for Z and ZH NPs. ZH NPs, in water, are less hydrated than Z NPs[45] (the time-averaged NP-water contacts of the ZH NP amount to 80% of the Z NP-water contacts). Water contacts are further reduced for the charged beads of the zwitterionic groups of ZH nanoparticles, as shown in Table 1, as a consequence of the ligand conformational change. Upon binding, it is the ZH NP that loses the largest number of water contacts, as shown in Table 1, coherently with the presence of a larger free energy barrier for binding (see also Figure 2).

**Table 1.** Number of contacts between NP beads and water beads, and between the charged beads of the zwitterionic groups and water beads. In parentheses the difference between the number of contacts in the bound and unbound state.

| NP | NP-water no NP-protein contact | NP-water during NP-protein contact | Zwitterionic group-water no NP-protein contact | Zwitterionic group-water during NP-protein contact |
|---|---|---|---|---|
| Z | 4909 ± 1 | 4889 ± 5 (-20) | 1542 ± 1 | 1533 ± 2 (-9) |
| ZH | 3950 ± 1 | 3821 ± 14 (-129) | 1015 ± 1 | 985 ± 4 (-30) |

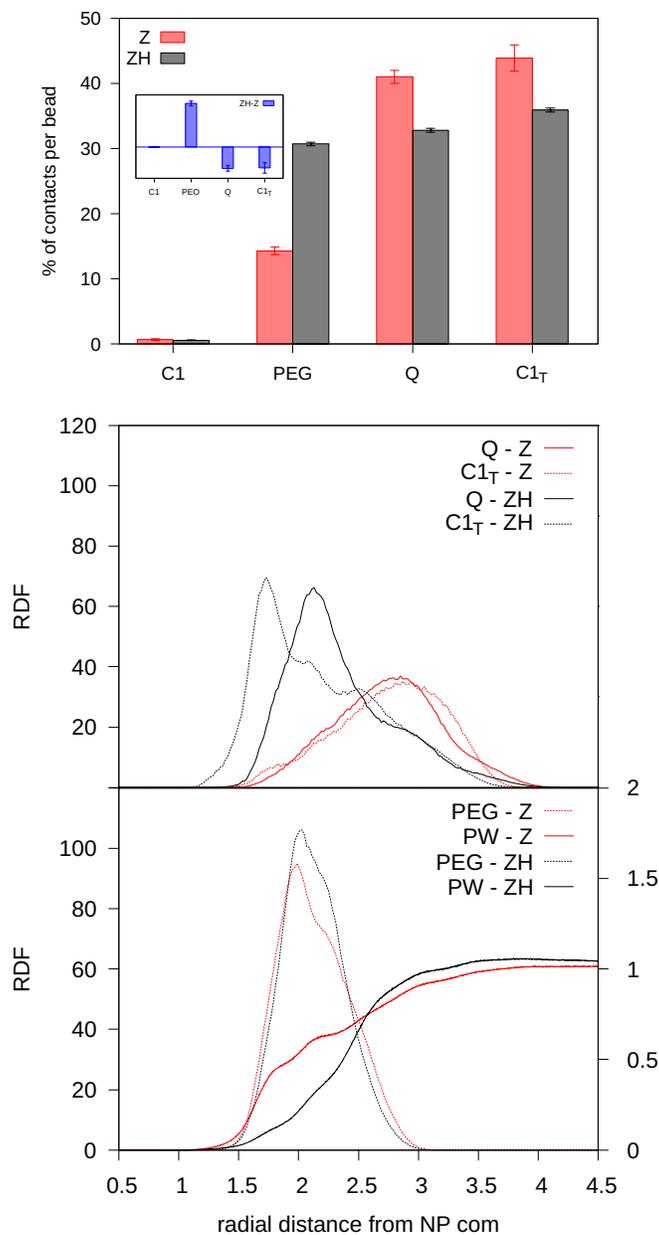

**Figure 4** – Top: percentage of NP-protein contacts involving different segments of the NP ligands: C1 refers to the hydrophobic groups next to the S atom, PEG refers to the 4 PEG monomers, Q refers to the charged groups of the zwitterion and $C1_T$ refers to the hydrophobic groups bound to the zwitterionic terminal. The inset shows the difference between the contact percentages of ZH and Z NPs, highlighting the increase of PEG-HSA contacts in the ZH case.

Bottom: radial distribution function of the different chemical groups composing the NP ligands. PW stands for polarizable water.

CONCLUSIONS

In this work we used coarse-grained molecular dynamics simulations with sub-molecular resolution to study the interaction of a monolayer-protected Au NP with the most abundant serum protein, HSA. We considered two types of NPs, functionalized by zwitterionic ligands with a different degree of hydrophobicity. Our simulations show that zwitterionic NPs have scarce propensity to form stable complexes with HSA, while more hydrophobic ligands interact more strongly with the protein – as measured in experiments by Moyano[13] *et al*. The excellent agreement with the experimental data allows us to interpret the experiments at the molecular level. The ligands terminated by hydrophobic groups interact more stably with the protein not by virtue of hydrophobic interactions, but because the hydrophobic moieties are folded towards the center of the NP and the PEG moieties are more exposed to the environment. NP-protein interactions, in this case, are determined by an increase of PEG-protein interactions, compatible with the formation of stable NP-protein complexes such as a hard protein corona.

Our data show that ligand conformation is as relevant as chemical affinity in determining protein-NP interactions. As a result, we propose that the design of protein-repellent NP functionalization should consider carefully the importance of both ligand conformation and ligand chemical composition. Computational models, also at coarse-grained level, are paramount for the prediction of ligand conformations relevant to the NP-protein interface, and we envision that they will contribute more in the future to quantify the relative weight of structural and chemical factors influencing NP-protein interactions.

ASSOCIATED CONTENT

**Supporting Information**. NP core and ligand parameterization; Human serum albumin parameterization; Simulation details; PEG ligands; Classification of protein residues; NP-protein contacts; Water-NP contacts; (PDF)

AUTHOR INFORMATION

The authors declare no competing financial interests.

ACKNOWLEDGMENT

Giulia Rossi acknowledges funding from the ERC Starting Grant BioMNP – 677513. LM acknowledges funding from INSERM. Calculations were carried out in part at CINES (grant A0040710138 to LM), in part at CINECA (grant HP10BOWTPR to GR). The authors thank Elisa Frezza for assistance in PCA analysis and fruitful discussions.

**TOC GRAPHICS**

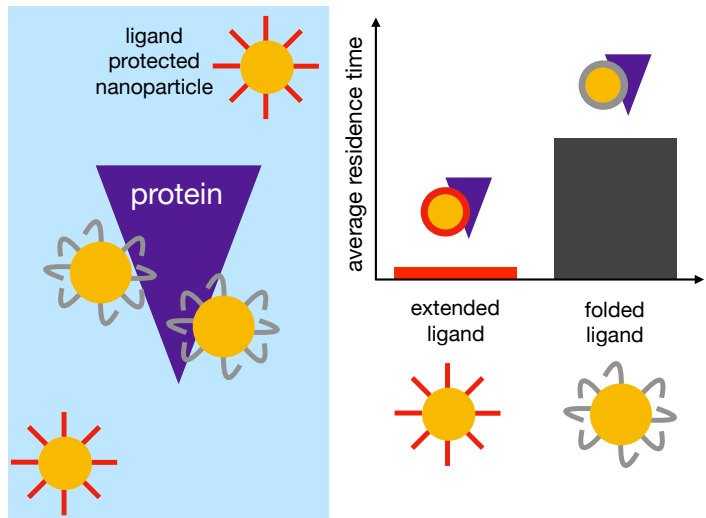

# Supporting Information for "Role of ligand conformation on nanoparticle-protein interactions"

*Federica Simonelli[1], Giulia Rossi[1]\* and Luca Monticelli[2]\**

[1] Physics Department, University of Genoa, Via Dodecaneso 33, 16146 Genoa, Italy
[2] MMSB, UMR 5086 CNRS / Universitè de Lyon, 7, Passage du Vercors, 69007 Lyon, France

## NP core and ligand parameterization

We considered functionalized gold nanoparticles (AuNPs) with a diameter of about 2 nm. The core of the AuNP consisted of 144 Au atoms. The structure of the core and the binding sites of the ligands on its surface were taken from Lopez-Acevedo et al.[1], while the elastic network connecting the Au and S atoms were described in Torchi[2] et al. The surface of the Au core has 60 binding sites for sulfur atoms. To functionalize each NP, we attached 60 identical ligands to the sulfur atoms. Two different types of ligands were considered, with different hydrophobicity. The less hydrophobic ligand (referred to as Z) contained only a betaine group. The more hydrophobic ligand, indicated in the following as ZH, featured a di-butane-sulfobetaine terminal. To parameterize the functionalized NPs we used the polarizable version of the coarse-grained Martini force field[3]. The ligands contained a hydrophobic chain of 9 $CH_2$ groups, which were modelled as 2 Martini C1 beads, a short chain of 4 poly-ethylene glycol monomers, represented by 4 Martini beads of type PEO (SN0 Martini type with redefined interactions according to Lee et al.[4] and modified angle interaction according to Bulacu et al.[5]) and the terminal zwitterionic group. For the PEG moieties, we also considered the latest MARTINI model by Grunewald et al.[6]

The parameterization of the terminal group was based on atomistic simulations by Ghobadi et al.[7]. We used the Martini force field to model an 8-monomer polymer containing a sulfobetaine group *per* monomer. Bead types for non-bonded interactions were chosen according to the Martini force field; the tetramethylammonium group was assigned a $Q_0$ type while the sulfonate ion was modeled as a $Q_{da}$ bead. The two beads were separated by a C1 bead. For the more hydrophobic ligand two butane chains were bound to the nitrogen atom in the tetramethylammonium ion and a C1 bead was used to model each butane chain. Bonded interactions were slightly modified to match atomistic simulations. In particular, we targeted the radial distribution function (RDF) of non-bonded $Q_{da}$ beads with respect to a $Q_0$ bead; we chose a $Q_0$ bead and computed the RDF of all $Q_{da}$ beads (except for the one bound to the $Q_0$ bead). The final set of parameters is summarized in Table T1.

**Table T1:** Bonded interaction parameters for Z and ZH ligands. Both ligands are shown in coarse grained representation under the interaction tables. S atoms in grey, C1 beads in red, PEO beads in orange and charged beads in yellow.

| Bonds | Length [nm] | Constant [kJ mol$^{-1}$ nm$^{-2}$] | Angles | Angle | Constant [kJ mol$^{-1}$ rad$^{-2}$] |
|---|---|---|---|---|---|
| S—C1 | 0.47 | 1250 | S—C1—C1 | 180 | 25 |
| C1—C1 | 0.47 | 1250 | C1—C1—PEO | 180 | 25 |
| C1—PEO | 0.45 | 5000 | C1—PEO—PEO | 180 | 25 |
| PEO—PEO | 0.33 | 7000 | PEO—PEO—PEO | 130 | CH:50 ReB:25 |
| PEO—$Q_0$ | 0.47 | 1250 | PEO—PEO—$Q_0$ | 180 | 25 |
| $Q_0$—C1 | 0.40 | 1250 | PEO—$Q_0$—C1 | 180 | 25 |
| $Q_0$—C1 (ZH side chains) | 0.47 | 1250 | PEO—$Q_0$—C1 (ZH side chains) | 90 | 25 |
| C1—$Q_{da}$ | 0.40 | 1250 | C1—$Q_0$—C1 | 180 | 25 |
| | | | $Q_0$—C1—$Q_{da}$ | 180 | 25 |

| Dihedrals | Angle | Constant [kJ mol$^{-1}$] | Multiplicity |
|---|---|---|---|
| PEO—PEO—PEO—PEO | 180 | 1.960 | 1 |
| PEO—PEO—PEO—PEO | 0 | 0.180 | 2 |
| PEO—PEO—PEO—PEO | 0 | 0.330 | 3 |
| PEO—PEO—PEO—PEO | 0 | 0.120 | 4 |

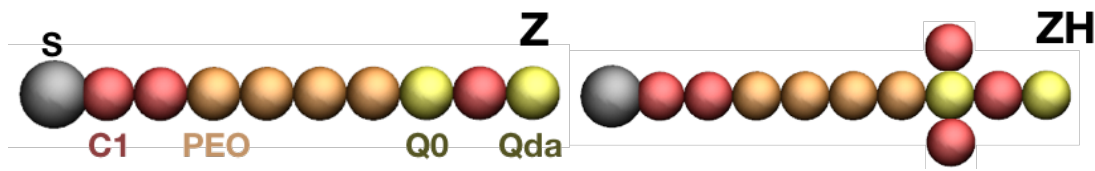

## Human serum albumin parameterization

Our model of human serum albumin (HSA) is based on the crystal structure of the protein reported in the PDB with code 1ao6. Only one of the two chains was kept for parameterization. The atomistic model was obtained with the GROMACS tool *pdb2gmx*. We used the atomistic Amberff99SB-ILDN[8] force field and the TIP3P water model. The protein was inserted in a simulation box of about 12x12x12 nm$^3$ with physiological salt concentration (150 mM KCl). A 1.5 μs run with velocity rescale thermostat (T=310 K, τ=1 ps) and Parrinello-Rahman barostat (isotropic, p=1 atm, τ=12 ps) was performed. Positions were saved every 100 ps to have enough sampling for principal component analysis (PCA).

The coarse-grained model of HSA is based on the extension to proteins of the Martini force field[9,10]. We tested both the elastic network and the Elnedyn[11] versions of the force field. We considered only the standard parameters for the elastic network while for the Elnedyn version we varied both backbone ($k_{BB}$) and elastic network ($k_{el}$) constants. We ran a 2 μs simulation for each coarse-grained model and compared the results with the atomistic model in terms of root mean square deviation (RMSD), root mean square fluctuations (RMSF), and PCA. To establish the best agreement with the atomistic model we use also the RMSIP index, that is the root meat squared inner product of the principal components (PCs). We consider only the first 20 PCs. The best set of parameters was: $k_{BB}$=90000 kJ mol$^{-1}$nm$^{-2}$, $k_{el}$=350 kJ mol$^{-1}$nm$^{-2}$, cutoff distance=0.9 nm.

## Simulation details

We performed molecular dynamics simulations in explicit polarizable water and at physiological salt concentration. For the zwitterionic NPs, we ran 20 simulations for each kind of NP. The initial configuration of these simulations was obtained by rotating the protein with respect to the NP which was kept in the same position in each initial configuration. One NP and one HSA were inserted in a simulation box of about 20x20x20 nm$^3$. Minimization and equilibration runs were performed before the simulation run. Each equilibration run was 20 ns long. Production runs were 3 μs long, each, for each NP and were performed with the velocity rescale thermostat[12] (T=310 K, τ=1 ps) and Parrinello-Rahman barostat[13] (isotropic, p=1 atm, τ=12 ps).

We performed 5 additional simulations with the PEG model by Grunewald, using the same setup as above and the same simulation parameters. No differences in the behavior of the 2 kinds of NPs were observed.

## PEG ligands

We performed 2 simulations of a single NP covered by 60 PEG ligands. Each ligand consisted of 9 monomers, i.e., the same number of beads as in the zwitterionic Z ligand. Each production run was 3 μs long. The same parameters and initial configuration setup were used in the simulations. The number of contacts between the PEGylated NP and HSA within 0.8 nm is shown as a function of time in figure S1 for the 2 simulation runs. Raw data are smoothed with a moving average algorithm (50 points in each averaging window). In both simulations, once the NP-protein contact was established, it remained stable throughout the simulation (3 μs).

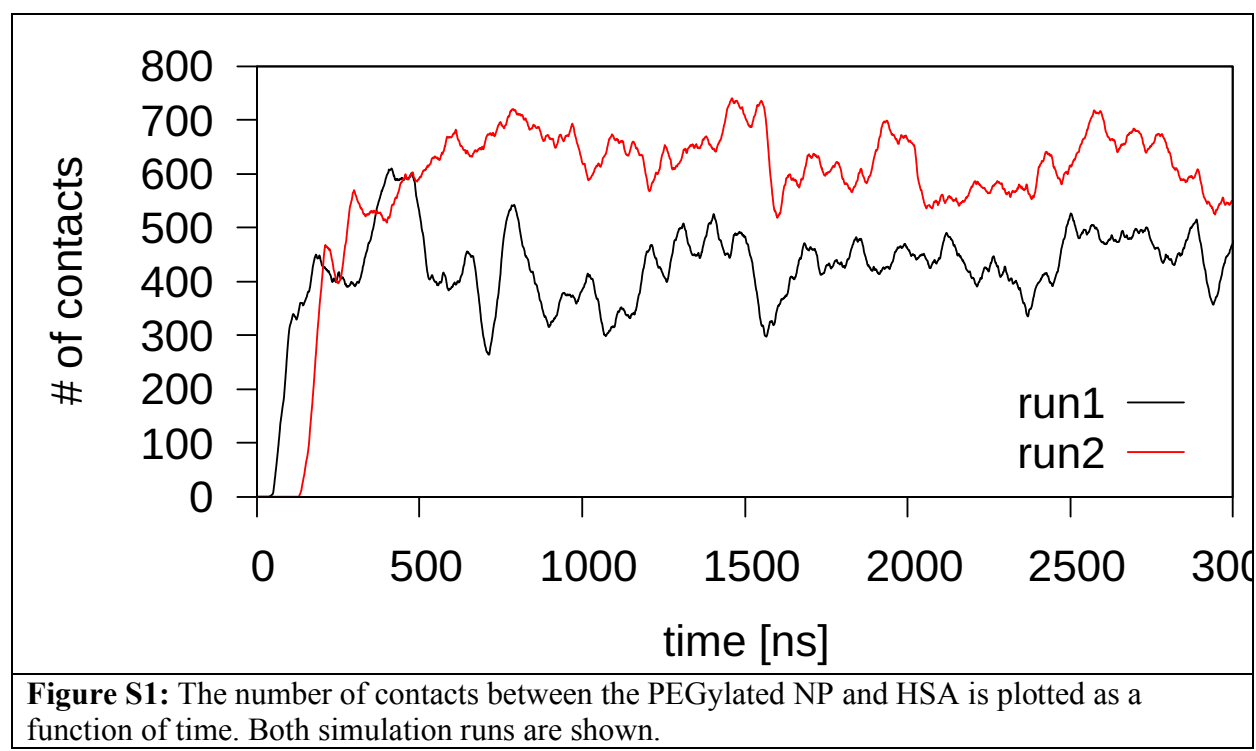

**Figure S1:** The number of contacts between the PEGylated NP and HSA is plotted as a function of time. Both simulation runs are shown.

## Classification of protein residues

Amino acids can be divided in charged, polar and hydrophobic (https://proteinstructures.com/Structure/Structure/amino-acids.html). Based on this classification, the residues in HSA are grouped to compute the number of contacts for different residue characters. Once divided by character, residues were grouped in surface and core residues according to their solvent accessible surface area (SASA). The SASA for each residue was computed from protein simulations in polarizable water using the GROMACS tool *gmx sasa*. A maximum value of SASA was then assigned to each amino acid according to Tien *et al.*[14]. The relative surface area, that is the ratio between the SASA and its maximum value for each residue, was used to discriminate between surface and core residues. If the ratio was below a threshold of 0.2, residues were considered as core residues, otherwise as surface residues. Only surface residues were considered in the calculation of the number of contacts.

# NP-protein contacts

The average number of contacts was computed as follow: first, the number of contacts between the NP and different groups of residues in the protein was computed as a function of time for each simulation; then an average over all simulations was computed, using the GROMACS tool *gmx analyze*. The average for each group was divided by the number of beads in the group. Final results were given as percentage of the total number of contacts.

# Water-NP contacts

The number of contacts between water and the 2 NPs were computed in two regimes: presence of NP-protein contacts and absence of NP-protein contacts. To determine the presence of a contact, a threshold of 0.8 nm was used (distance between NP and protein beads). The average number of contacts was computed considering all the stretches of contact/non-contact for the 20 simulations with the GROMACS tool *gmx analyze*. Only the W bead (the only one with Van der Waals interactions) of polarizable water was considered for contacts.